\documentclass[12pt,aps,a4paper,nofootinbib]{revtex4}
\usepackage{amsmath}
\usepackage{graphicx}
\usepackage[usenames,dvipsnames]{color}

\newcommand{\re}{\ref}
\newcommand{\be}{\begin{equation}}
\newcommand{\ee}{\end{equation}}

\newcommand{\la}{\label}
\newcommand{\ber}{\begin{eqnarray}}
\newcommand{\eer}{\end{eqnarray}}

\newcommand{\lcdot}{\boldsymbol{\cdot}}
\begin{document}

\title{On the Method of Projection onto Localized States\\ for Describing Reactions and Clustering}

\author{V.D. Efros\footnote{Email: v.efros@mererand.com}}

\affiliation{National Research Centre "Kurchatov Institute", Moscow, Russia}


\begin{abstract}
The method of projection onto localized states  intended for describing reactions and
clustering is examined. In particular, the method offers a simple extension of the no-core shell model to
continuum- and clustered bound states.
A critical analysis of a previous realization of the method is provided. 
A way to minimize
the number of basis functions required for convergence is given. An easy way to obtain exactly
unitary and symmetric $S$ matrix from calculations is pointed out. 
The method is tested in one- and two-channel $np$ scattering problems
with the Daejeon $NN$ interaction. It is shown that excellent convergence can be reached even when only
oscillator states with low number of excitation quanta are retained.
\end{abstract}

\bigskip

\maketitle

\section{Introduction}

Over the last decades,
various approaches to ab initio description of nuclear reactions have been developed, see, e.g., 
Refs.~\cite{john,sha} for a review.
 For systems with ${\rm A}>4$, apart from the integral transform studies, 
realistic calculations are typically performed using an approach~\cite{qua,nav,gys} that combines 
the no-core shell model (NCSM)  with the resonating group
method or incorporates  the description of continuum into NCSM.
The calculations are technically involved due to the need to solve  the set of integro-differential equations
whose kernels are rather complicated and
include tedious antisymmetrization between fragments participating in the reaction. In Ref.~\cite{efr},
a simpler method to extend NCSM on nuclear reactions has been given. It avoids the mentioned antisymmetrization
and, like the NCSM, requires solving only linear equations. 

In  one-channel two-body
problems with model potentials this method yields quite satisfactory results \cite{efr,efr1}. 
To test it further, 
Refs.~\cite{sha,sha1} applied it 
to calculate phase shifts, mixing parameters and wave functions for one- and two-channel
$np$ scattering with the 
Daejeon16  $NN$ interaction \cite{da}. The approach has also been extended~\cite{sha} to treat
cluster bound states and 
resonances. Successful results have been obtained in the latter problems  with both the Daejeon16 $NN$ interaction 
and model potentials. However, in  scattering calculations,
instabilities were encountered and  convergence of the results proved to be incomplete in the two-channel case.

In the present work, we re-examine the issue. The equations to be solved \cite{efr,zhuefr} are of the form
$\int d\tau{\bar \chi}_k^*(H-E)\Psi=0$, where $\Psi$ is a trial reaction wave function 
and ${\bar \chi}_k$ is a set of localized functions. 
 We show that  for specific choices of ${\bar \chi}_k$ and ansatz for $\Psi$,
such 
as in Refs.~\cite{sha,sha1},
the resulting $\Psi$ is always singular for some energies, strongly differing from the true reaction wave function in 
the vicinities of these energies.

Whereas with   ${\bar \chi}_k$ and ansatz for $\Psi$ chosen
differently, highly stable results are obtained in the present work. They 
are very accurate 
even when only low-excitation
oscillator states are retained in the calculations, demonstrating the efficacy of the method.

A way is also suggested to optimize nonlinear parameters in the trial functions $\Psi$.
An easy way is pointed out to obtain exactly unitary $S$ matrix 
from a dynamical calculation. All this have
contributed to the quality of the present scattering results. 

The method is outlined in the next section. 
In Sec. III A, a description of the aforementioned abnormal solutions is provided.
In Sec. III B, a way for obtaining exactly unitary $S$ matrix from a dynamic calculation is presented.
In Sec. III C, a suitable way for choosing non-linear parameters that enter trial continuum wave functions
is given. In Sec. IV, the method is tested via the calculation  
of phase shifts and mixing parameters of $np$ scattering with the Daejeon16 $NN$ interaction.

\section{The method}

To proceed, let us outline the approach we shall employ. For brevity, we will refer to wave functions,
or their components, 
having finite norms as localized. We
consider two-fragment reactions and use the ansatz 
of the form  
\be\Psi_i=\sum_{k=1}^Pc_k^i\chi_k+\sum_{k=P+1}^{P+Q}c_k^i\psi_{q}^{(1)}+\psi_i^{(0)}\la{psi}\ee
to approximate continuum-spectrum wave functions.
Here, $\chi_k$ are localized functions (LF), while $\psi_i^{(0)}$ and $\psi_{q}^{(1)}$ are the 
non-localized "channel" functions.
The latter are antisymmetrized products, coupled to a total momentum, of 
bound-state wave functions of fragments and a function of their relative motion. There are two
types of relative-motion functions, distinguished
by the superscripts $(0)$ and $(1)$. In the second sum of Eq. (\re{psi}), $q=k-P$.
 Correspondingly, $q$ takes the values $1,\ldots,Q$, while $i$ is one of
these values.   The number of open channels is $Q$. (Closed channels 
could also be incorporated.)

Within the interior region, LF $\chi_k$ and
the channel functions
$\psi_i^{(0)}$ and $\psi_q^{(1)}$  contribute to the wave functions~$\Psi_i$.
  However, at larger distances,
$\Psi_i$ are formed exclusively by the  functions $\psi_i^{(0)}$ and $\psi_q^{(1)}$.
Beyond the interior region, 
the radial parts of the relative-motion  functions entering $\psi_i^{(0)}$ and~$\psi_q^{(1)}$
describe motion in
the average interfragment Coulomb
potential, or corresponding free motion.
The relative-motion functions are taken to be regular at small distances. 
The center-of-mass motion factor is omitted in Eq.~(\re{psi}). 
The $S$~matrix is
given by the $c_k^i$ coefficients entering the second sum in Eq.~(\re{psi}).

The functions $\Psi_i$ of Eq. (\re{psi})  form should approximate 
true  continuum wave functions, and we need to find the $c_k^i$ coefficients ensuring this.
The following procedure is applied for this purpose. 
Consider the functions $(H-E)\Psi_i$, where $H$ is the Hamiltonian of the system and $E$ is its total energy.
It is straightforward to verify, see also \cite{efr}, that these functions  are localized, i.e. the norm
$\parallel (H-E)\Psi_i\parallel$ is finite. 
(Note that this property does not hold 
for
generalizations
of the ansatz (\re{psi}) that include open channels with more than two fragments.) 
We should, at least, impose the condition 
\be\parallel (H-E)\Psi_i\parallel\simeq0.\la{co}\ee
Let ${\bar\chi}_j$, $j\ge1$, be linearly independent localized functions that form an $L^2$-complete set in the 
space of localized functions. For convenience, this set can be taken as orthonormalized.
Let us denote the scalar products\footnote{We shall denote scalar products as 
$/\alpha/\lcdot/\beta/$, where $/\alpha/$ and $/\beta/$ are state vectors. (Thus, $/\alpha/$ is not a
bra vector, see \cite{efri}.)
Apart from scalar products, we deal everywhere with wave functions.  Then, when passing to 
scalar products, we use akin notation for state vectors and corresponding wave functions, such as $/\Psi/$ 
and $\Psi$.} 
 $/{\bar\chi}_j/\lcdot(H-E)/\Psi_i/$ as~$d_j$.  One has $\parallel (H-E)\Psi_i\parallel^2=\sum_{j=1}^\infty |d_j|^2$. 
Therefore, $d_j\simeq0$ for sufficiently large $j$.
Then, to meet requirement (\re{co}),
one can set $d_j=0$ for $j\le P+Q$, i.e.,
\cite{efr,zhuefr} 
\be /{\bar\chi}_j/\lcdot(H-E)/\Psi_i/=0\quad{\rm at}\quad j=1,\ldots, P+Q\la{eqs}.\ee 
This forms a system of $P+Q$ linear equations for $P+Q$  coefficients $c_k^i$.
(A reasonable choice in certain cases is to set 
${\bar\chi}_j=\chi_j$ for $j\le P$.)

 It seems plausble that the Eq.~(\re{psi})
wave functions with the~$c_k^i$
coefficients obtained from Eqs.~(\re{eqs}) well approximate  true continuum-spectrum wave functions 
when the number $P$ of LF retained in the trial function (\re{psi}) is sufficiently large.
In fact, if the functions ${\bar\chi}_j$ and  LF $\chi_k$
appearing in Eq.~(\re{psi}) are defined in a finite-dimensional space, there
exist peculiar cases when this is not correct, see below.

The system (\re{eqs}) of equations involves  matrix elements of the following two types:
\be /{\bar\chi}_{j}/\lcdot(H-E)/\chi_{k}/\quad{\rm and}
\quad /{\bar\chi}_{j}/\lcdot(H-E)/\psi_{i,q}^{(0),(1)}/.\la{me}\ee
The matrix elements of the first type are standard within the framework of the ncsm.
The calculation procedure for the matrix elements of the second type is presented in Ref.~\cite{efr}.
When calculating these matrix elements,  
the interfragment antisymmetrization is not required
\cite{efr,zhuefr}. 

In the case of the  equations (see, e.g., \cite{dem})
pertaining to the Hulth\'en-Kohn variational method, where   
 the ansatz~(\re{psi})  is also used,
additional  matrix elements of the type
\be /\psi_r^{(0),(1)}/\lcdot(H-E)/\psi_s^{(0),(1)}/\la{me1}\ee
arise. Moreover, the scope of applicability of the Hulth\'en-Kohn variational principle  is limited
in the
many-body case, see \cite{efr,dem}.

Projection equations of Eqs. (\re{eqs}) type are also useful for calculating two-cluster bound states, such as $^7$Li or 
$^7$Be \cite{sha}. One writes the bound-state wave function as
\be\Psi=\sum_{k=1}^Pc_k\chi_k+\sum_{k=P+1}^{P+Q}c_k\psi_{q},\quad q=k-P,\la{psib}\ee
where $\chi_k$ are appropriate ncsm superpositions of Slater determinants and the terms $\psi_{q}$ are 
responsible for the description of clusterized tails of the nucleus. 
These terms are of similar structure
as the channel functions in Eq. (\re{psi}).  Beyond the interior region, 
the functions of relative motion entering $\psi_{q}$ represent motion 
in an average Coulomb potential
with negative relative-motion energy exactly corresponding to the total energy of a level.

Now Eqs. (\re{eqs}) with $\Psi_i$ replaced
by $\Psi$ and with $E$ being the energy of a level,
\be/{\bar\chi}_j/\lcdot(H-E)/\Psi/=0\quad{\rm at}\quad j=1,\ldots, P+Q,\la{eqsb}\ee
are homogenous equations for the $c_k$ coefficients. They are of the form 
\be\sum_{k=1}^{P+Q}a_{jk}(E)c_k=0.\la{heq}\ee 
The energy $E$ is to be determined in advance by solving the eigenvalue problem 
$\det a(E)=0$. The latter problem 
can be solved using the Newton method. 

Here, an important feature is that, unlike $\Psi$ itself, the quantity $(H-E)\Psi$ is short-ranged
to a good approximation.\footnote{The contribution to $(H-E)\psi_{q}$
from the residual Coulomb interaction
between clusters
 is not short-ranged, but this contribution is small and can be treated as a 
perturbation. This residual Coulomb interaction is not explicitly accounted for in
the expressions (\re{psi}) and (\re{psib}).} Therefore, in the present approach one can expect faster
convergence than in the case of a variational calculation with no 
cluster components included in a trial function. And if one includes such components in a variational 
calculation then one needs to evaluate matrix elements of Eq. (\re{me1}) type in addition.
In Ref. \cite{sha}
it has been shown that for the deutron case convergence in this approach is indeed much faster then in 
a variational calculation using the oscillator basis.   

Now let us proceed to the present findings.

\section {Additional issues}

Subsection A below is relevant, in particular, to the calculations of the preceding work,
but not to those 
of the next section.  In contrast, the outcomes of subsections~B and C
 are used in the next section's
calculations.

\subsection{Abnormal solutions of the equations}

This subsection studies some of the cases, where  eigenfunctions of a Hamiltonian
in a finite-dimensional space
are taken   as LF $\chi_k$ in Eq. (\re{psi}). 
Consider the following choice of~$\chi_k$ in Eq. (\re{psi}) and ${\bar\chi}_j$ in  Eqs. (\re{eqs})
(recall that in Eq. (\re{psi}) $k=1,\ldots,P$ and in Eqs. (\re{eqs}) $j=1,\ldots,P+Q$):
 
a) Let us denote  a finite-dimensional space of LFs as ${\cal N}$. Normalized eigenfunctions 
of  a Hamiltonian in~${\cal N}$
 are taken as LF $\chi_k$ in Eq. (\re{psi}). 

b) In  Eqs. (\re{eqs}), for $j\le P$, it is set that ${\bar\chi}_j=\chi_j$.

c) The remaining
$Q$ functions  ${\bar\chi}_j$ in  Eqs. (\re{eqs}), namely those with $j=P+1,\ldots,P+Q$,  are taken to 
belong {\it just to the space in which
the aforementioned eigenfunctions $\chi_k$ are defined}. This condition is determinative.

Such a choice of $\chi_k$ and ${\bar\chi}_j$ has been made in Ref. \cite{sha}, where $\cal N$ was the
space formed by oscillator
functions with excitations up to certain values.

We will show that, with $\chi_k$ and ${\bar\chi}_j$ chosen in the described way,  
solutions to Eqs. (\re{eqs}) exhibit singularities for some sets of energies.

To this aim, we will transform the functions ${\bar\chi}_j$ with $j=P+1,\ldots,P+Q$. We represent these
functions as linear combinations of  
eigenfunctions of a Hamiltonian in the 
space~$\cal N$.  This can be done since in the case we consider
functions ${\bar\chi}_j$ belong to this space. According to conditions a) and b) above,
some of the mentioned eigenfunctions are LF $\chi_{k}$ that appear in Eq.~(\re{psi}).
We will remove the contributions of the latter eigenfunctions from those linear combinations.
We denote the functions arising as a result as ${\widetilde\chi}_j$. 
We replace the equations~(\re{eqs}) for \mbox{$j=P+1,\ldots,P+Q$} with the equations 
\mbox{$/{\widetilde\chi}_j/\lcdot(H-E)/\Psi_i/=0$}. Thus, the resulting  set of equations consists
of these new equations plus the original equations \mbox{$/\chi_j/\lcdot(H-E)/\Psi_i/=0$}
for $j=1,\ldots,P$.

Two sets of functions, namely the original set ${\bar\chi}_j$ and the new set comprising
 $\chi_{j}$ for $j\le P$ and 
${\widetilde\chi}_j$ for $j>P$
can be obtained from each other through linear transformations. Consequently,
the same refers to the original set (\re{eqs}) of equations  and the newly listed set of equations.
Therefore, the solutions $\{c_k^i\}$ of these two sets
of equations are the same. The determinants of the left-hand side matrices  
for these two sets of equations are also equal.

The equations of the latter set look as follows,
\begin{subequations}\la{6}
\be (E_j-E)c_j^i+\sum_{{k=P+1}}^{P+Q}a_{jk}c_{k}^i=b_j^i\quad{\rm at}\quad j=1,\ldots,P,\la{eq1}\ee
\be \sum_{k=P+1}^{P+Q}a_{jk}c_{k}^i=b_j^i\quad{\rm at}\quad j=P+1,\ldots,P+Q.\la{eq2}\ee
\end{subequations}
In Eqs. (\re{eq1}), $E_j$ are the corresponding eigenvalues, 
 \[a_{jk}=/\chi_j/\lcdot(H-E)/\psi^{(1)}_{k-P}/,\quad 
b_j^i=-/\chi_j/\lcdot(H-E)/\psi^{(0)}_{i}/.\] In Eqs. (\re{eq2}),
\[a_{jk}=/{\widetilde \chi}_j/\lcdot(H-E)/\psi^{(1)}_{k-P}/,\quad
b_j^i=-/{\widetilde \chi}_j/\lcdot(H-E)/\psi^{(0)}_{i}/.\]
Here, the present-case relations $/\chi_j/\lcdot/\chi_{k}/=\delta_{jk}$
and $/{\widetilde \chi}_j/\lcdot(H-E)/\chi_{k}/=0$
are taken into account.  
Eqs. (\re{eq2}) determine the $S$ matrix. 

When the energy $E$ coincides with one of the $E_j$ values, the determinant of the system of equations (\re{6}) 
vanishes because the $j$-th column, where $j\le P$, of the matrix on its left-hand side then consists entirely of zeros.
Furthermore, if for the corresponding $j$ value,   
the relation $\sum_{k=P+1}^{P+Q}a_{jk}c_{k}^i-b_j^i\ne0$  holds true at $E=E_j$,
then
the  coefficient $c_j^i(E)$ exhibits a pole
at $E=E_j$, as seen from Eqs. (\re{eq1}).

Normally, the above relation is valid. Indeed, at $E=E_j$, let us consider the following system of equations 
for the $c_{k}^i$ coefficients:
$\sum_{k=P+1}^{P+Q}a_{jk}c_{k}^i=b_j^i$, where one of the equations is the
relation just opposite to that listed above ($j\le P$) and the other equations are Eqs.~(\re{eq2}) ($j>P$).
If the relation we consider would not be
true, then this system of equations would have a solution. 
But this is a system of $Q+1$ linear equations for $Q$ variables which 
might have a solution only occasionally, when the first mentioned equation is a consequence of Eqs.~(\re{eq2}).

As seen from Eqs. (\re{6}), at $E$ tending to $E_j$ all the $c_k^i$ coefficients except for that with $k=j$
normally remain finite. At $E\rightarrow E_j$ this leads to the domination in $\Psi_i$ at finite distances 
 of the contribution of $\chi_j$,  due to the $(E-E_j)^{-1}$ pole.

Thus, at the choice of $\chi_k$ and ${\bar\chi}_j$  described here, 
the wave functions (\re{psi}) calculated from Eqs. (\re{eqs}) normally
cannot properly approximate true continuum wave functions 
in the vicinities of the
eigenvalues $E_j$, irrespective to the number  
of   LFs retained in Eq. (\re{psi}).
We shall not study features of the solutions of Eqs. (\re{6}) further since 
a different natural choice of the basis of LFs employed in the next
section provides accurate and stable results at any energies.\footnote{This behavior stands in contrast 
to the case of the Hulth\'en-Kohn system of equations. 
The left-hand side matrix of these 
equations is Hermitian and, consequently, at energies equal to eigenvalues of this matrix,
singularities arise at any choices of LFs in Eq. (\re{psi}), see Refs.~\cite{sch,nut}.} 

The anomalies described above should have a relation to numerical and convergence problems in the
calculations of Refs. \cite{sha,sha1}. However, the  determinant of the equivalent system
of equations computed in Ref. \cite{sha1} in the vicinity of the first excited state, which should
behave in the same way as above, does not exhibit this behavior.

\subsection{Making $S$ matrix unitary}

For $S$ matrix derivable from
Eqs. (\re{eqs})
the symmetry and unitarity properties fulfill only approximately. 
But in order to extract $S$ matrix from experiment, one uses its parametrizations implying
that these properties fulfill exactly. Therefore, for comparison with the corresponding parameters deduced
from experiment,
one needs to get exactly  unitary $S$ matrix from a calculation.

In Ref. \cite{sha},
the channel functions 
$\psi_i^{(0)}$ and $\psi_q^{(1)}$ in Eq. (\re{psi}) were chosen such that the coefficients $c_k^i$ with $k>P$   
were elements of $S$ matrix. 
The symmetry relation $S_{\alpha\beta}=S_{\beta\alpha}$ was imposed 
but  it is not possible to obtain an exactly unitary $S$ matrix in this way.

The simplest possibility to attain this goal is the following.
In Eq. (\re{psi}), let us choose
$\psi_i^{(0)}$ and $\psi_q^{(1)}$ such 
that the   coefficients $c_k^i$ with \mbox{$k>P$}
are elements of $K$ matrix. The true $K$~matrix is symmetric and real. 
The $K$~matrix obtained from Eqs. (\re{eqs}) is real automatically. Its symmetrization can be done directly.    
(If $K_{\alpha\beta}\simeq K_{\beta\alpha}$  one can adopt the average of these two
values. Otherwise, one can employ the one of them  
for which convergence is faster.) The
$S$~matrix sought-for is  to be expressed
in terms of real and symmetric $K$ matrix thus obtained, Eq. (\re{s}) below.  Then the resulting $S$ matrix
 will automatically be 
exactly unitary and symmetric. 

\subsection{Optimization of values of parameters}

The functions of relative motion of fragments entering the channel functions  
from Eq.~(\re{psi})
include parameters.  Namely, considering the formulation that employs $K$ matrix, 
 let us write the radial part of the relative-motion function entering $\psi_{q}^{(1)}$ 
 as ${\rm const}\cdot{\tilde G}_l(k,r)/r$, where $k$ is the wave number,
$r$ is the distance between the centers of mass of fragments, and
$l$ is the orbital momentum.
The function ${\tilde{G}_l}$
is obtained from the irregular Coulomb function $G_l$ by means of a regularization at small distances. One
can set ${\tilde G}_l(k,r)=g_l(r)G_l(k,r)$, where the factor $g_l$ ensures the regularization.
Possible choices of $g_l$ are given by:
\begin{subequations}\la{gl}
\be g_l(r)=\left[\tanh\left(r/r_{cut}^l\right)\right]^{2l+1},\la{fact1}\ee
\be g_l(r)=\left[1-\exp\left(-r/r_{cut}^l\right)\right]^{2l+3},\la{fact2}\ee
\end{subequations}
where $r_{cut}^l$ are parameters.
Let
$T_{rel}$ and $E_{rel}$ be, respectively,  the operator of relative kinetic energy  
and energy of relative motion of fragments.  
It can be verified that, with the choices~(\re{gl}), the
quantity  $(T_{rel}-E_{rel})\psi_{q}^{(1)}$ 
is regular at $r$ tending to zero.
In the case of the choice (\re{fact1}) 
the function ${\tilde G}_l(k,r)/r$ behaves  as
$r^{l}$ as $r$ tends to zero, i.e., like $F_l(k.r)/r$, $F_l$ being the regular Coulomb function.
In the case of the choice (\re{fact2}) the function ${\tilde G}_l(k,r)/r$ behaves  as
$r^{l+2}$ as $r$ tends to zero.

Let us address the issue of choosing values of parameters like $r_{cut}^l$ in Eqs. (\re {gl}).
If they are chosen improperly, then
a large number of LFs  
in Eq.~(\re{psi}) is required
 to compensate the corresponding "wrong" behavior at small distances of the channel functions. 
In the many-body case, this number would be excessively large and achieving convergence would be problematic.
 Thus,    
we want to find such values of the parameters  that 
the calculated 
$S$~matrix becomes close
to the true one already at a sufficiently small
 number~$P$ of  LFs retained in Eq. (\re{psi}).

To this aim, we will proceed as follows. 
Consider the quantities
$ |\nabla c_k^i/c_k^i|$ with $k>P$, where the gradients are taken with respect to the 
parameters. For a given~$P$, we will adopt
the  values of the parameters that minimize these quantities as the optimal ones. More practical is to use
finite difference analogues of the gradients, see the next section.

This procedure may be explained as follows.
If a sufficiently large number of LFs were retained in a calculation,
the $c_k^i$ coefficients with $k>P$ would presumably cease to depend on the
parameter values. With a smaller number of retained LFs, we approach this regime by
seeking parameter values  that minimize this dependence. The procedure is not
universal - for example, a minimum may not always exist - but in the calculations below it has proven to be useful.

\section{Calculations of ${\it  \boldsymbol{np}}$ scattering}

\subsection{General description}

In the calculations below, LF $\chi_k$ from Eq. (\re{psi}) are taken as the oscillator functions.  
The purpose of these calculations is as follows. In the many-body case, using the set 
of the Lanczos basis functions as
LF $\chi_k$ in Eq. (\re{psi}) is advantageous, see also \cite{efr}. Such LF are represented as
combinations of oscillator Slater determinants, and, for
practical reasons, these combinations
should not
include determinants with too high excitations. Accordingly, like
Refs.~\cite{sha,sha1},
we must determine  whether the calculated reaction yields remain close to the true ones 
when only oscillator functions with sufficiently low numbers of excitation quanta are retained.

In our case, the LF $\chi_k$ are
products
of oscillator functions and two-nucleon spin functions
coupled  to total angular momentum:
\be\chi_k= R_{nl}(r){\cal Y}_{JM}^{ lS},\la{bas}\ee
where $r$ is the distance between nucleons, $n$ is the oscillator radial quantum number, 
$l$, as above, is the orbital momentum,
$S=0$ or 1 is the spin of the two-nucleon system, $J$~and~$M$ are the total angular momentum 
and its projection, ${\cal Y}_{JM}^{lS}$ is the standard spin-angular function, and  $R_{nl}$ is
the radial oscillator function. One has $n=n(k)$ and \mbox{$l=l(k)$}. The oscillator
energy is $\hbar\omega(2n+l+3/2)$, where $\omega$ is the oscillator frequency. 
The Daejeon16 $NN$ interaction~\cite{da} we employ is constructed
in the form of matrix elements between oscillator functions~(\re{bas}) with
$\hbar\omega=25$ MeV. In the present calculations,
we use the functions~(\re{bas})  with just this $\omega$ value.
In the many-body case,
the possibility to optimize a calculation with respect to~$\omega$ value is restricted by the condition
that this value should be the same for LFs~$\chi_k$ and for wave functions
of fragments entering the channel functions.

One has
\be R_{nl}(r)=r_0^{-3/2}[2n!/\Gamma(n+l+3/2)]^{1/2}x^l\exp(-x^2/2)L_n^{l+1/2}(x^2).\la{r}\ee
Here $L_n^\alpha(z)$ are Laguerre polynomials and  
$x=r/r_0$, where \mbox{$r_0=(\hbar/\mu\omega)^{1/2}$}  is the oscillator radius with  $\mu$ being the reduced mass of 
the $np$ system: \mbox{$1/\mu=1/M_n+1/M_p$}, where $M_n$ and~$M_p$ are the neutron and proton masses. We adopt the value
\mbox{$\hbar^2/(2\mu)=41.47107$}~MeV$\cdot$fm$^2$. The functions $R_{nl}$ 
are normalized to unity and the basis functions
(\re{bas}) are thus orthonormalized. 

The functions ${\bar \chi}_j$ from Eqs. (\re{eqs}) are of the same form (\re{bas}) with $R_{nl}$ of Eq.  (\re{r}).

Let us list the matrix elements (\re{me}) for the present case. The first-type matrix elements are as follows.
Write $H=T+V$, where $T$ and $V$ are kinetic
and potential energy. The matrix elements of $T$ between LFs of Eq. (\re{bas}) form have the structure
$T_{jk}=T_{nn'}^l\delta_{ll'}$, where $n=n(j)$, $l=l(j)$,
 $n'=n'(k)$, and $l'=l'(k)$. 
These matrix elements are equal to zero
if $|n-n'|>1$. The non-zero matrix elements are given by
\[T_{nn}^l=\frac{\hbar\omega}{2}\left(2n+l+\frac{3}{2}\right),\quad T_{n-1,n}^l=T_{n,n-1}^l=
\frac{\hbar\omega}{2} \left[n\left(n+l+\frac{1}{2}\right)\right]^{1/2}.\]
The matrix elements of $V$ between LFs of Eq. (\re{bas}) form are denoted as $V_{nl,n'l'}^{JS}\equiv V_{jk}$,
and  are taken from Ref. \cite{da}.

The second-type matrix elements from Eq. (\re{me}) can be calculated, in particular, as follows.
In the present case, the channel functions are given by: $\psi_{i}^{(0)}=j_l(kr){\cal Y}_{JM}^{lS}$, where $l=l(i)$ and 
\mbox{$\psi_{q}^{(1)}=g_l(r)n_l(kr){\cal Y}_{JM}^{lS}$}, where $l=l(q)$. 
Here, $j_l(z)$ and $n_l(z)$
are the spherical Bessel and Neumann functions and, as above, $\hbar k=(2\mu E)^{1/2}$ is 
the relative momentum of colliding particles. 
The matrix elements  are given by:
\begin{subequations}\la{7} 
\be/\chi_j/\lcdot(H-E)/\psi_{i}^{(0)}/=/\chi_j/\lcdot V/\psi_{i}^{(0)}/=
\sum_{n'=0}^{{n_0}}V_{nl,n'l'}^{JS}\int_0^\infty r^2dr
R_{n'l'}j_{l'},\la{mee}\ee
\begin{eqnarray}/\chi_j/\lcdot(H-E)/\psi_{q}^{(1)}/=
-\frac{\hbar^2}{2\mu}\int_0^\infty r^2drR_{nl}[g_{l}^{''}n_l+2g_{l}'(n_{l}'+n_l/r)]\nonumber\\
{}+\sum_{n'=0}^{{n_0}}V_{nl,n'l'}^{JS}\int_0^\infty r^2dr R_{n'l'}g_{l'}n_{l'}.\la{me2}\end{eqnarray}
\end{subequations}
In these relations, $n=n(j)$, $l=l(j)$, $l'=l'(i)$ or $l'(q)$, and $g_{l}'\equiv dg_{l}/dr$, etc. 
The matrix elements~$V_{nl,n'l'}^{JS}$
from Ref. \cite{da} are different from zero only if both $2n+l$ and $2n'+l'$ do not exceed 60 or 61. Correspondingly,
 in the above relations ${n_0}=[(61-l')/2].$

All the terms in the sums over $n'$ in Eqs. (\re{7}) are included in the calculations. In the 
many-body case, only limited numbers of terms are to be retained in such type sums. For the present case, restricting
these sums to a reasonably small numbers of terms does not significantly alter the final results. For instance,
in the $^1S_0$ scattering case, truncating the sums at the value $n'=8$ changes the phase shift by 2, 3, and 6\%
at energies 2, 24, and 50 MeV, respectively. These changes can be made even less due to the modification 
of the standard way employed in Eqs. ({\re{7}) of approximating the functions $j_{l'}$ and $g_{l'}n_{l'}$ with the
functions~$R_{n'l'}$, see Ref. \cite{efr}.

\subsection{Single-channel scattering}

Here, scattering in the $^1S_0$ channel is considered. In Eq. (\re{psi}) with $Q=1$ we set 
\mbox{$\psi^{(0)}_1=j_0(kr){\cal Y}_{00}^{00}$} 
and $\psi^{(1)}_1=g_0(r)n_0(kr){\cal Y}_{00}^{00}$. The factors $g_0(r)$ are as in  Eqs.~(\re{gl})
for $l=0$. One has
$c_{P+1}=\tan\delta$, where $\delta$ is the phase shift. The LF $\chi_k$ are $R_{n0}{\cal Y}_{00}^{00}$, 
where $n=k-1=0,\ldots,P-1$. The functions
${\bar \chi}_j$ are of the same form with $n=j-1=0,\ldots,P$. The total number $P+1$ of 
retained equations, i.e., of the functions 
${\bar \chi}_j$, will be denoted as~$N$. The maximum number of excitation quanta is $2(P-1)=2N-4$.

 The parameter $r_{cut}^0$ entering $g_0(r)$ is optimized by minimizing the 
quantity
  \be\left|\frac{\delta(r_{cut}^0+0.25\, {\rm fm})-\delta(r_{cut}^0-0.25\, {\rm fm})}{\delta(r_{cut}^0)}\right|\la{opt}\ee 
  on the grid of $r_{cut}^0$ values with the step 0.25 fm over the interval 1 fm $\le r_{cut}^0\le$ 4 fm.
The minimizations have been
performed for energies $E=0.05$, 1, 2, 10, 24, and 50~MeV. 
 
Let us consider first the results obtained with the choice (\re{fact1}): $g_0(r)=\tanh(r/r_{cut}^0)$. 
The minimization procedure has been done at $N=8$.  
 The $r_{cut}^0$ values thus obtained are 
2.25 fm at $E=0.05$, 2, and 10~MeV; 2.5 fm at $E=1$ and 24 MeV, and 2.75 fm at $E=50$ MeV. For simplicity,
the same value of $r_{cut}^0$
is adopted for all the $E$ and $N$ values in the subsequent calculations: $r_{cut}^0=2.25$~fm.

Fig. 1 shows the results obtained for the phase shift $\delta$.
The top panel refers to the full energy range, whereas the bottom panel highlights the low-energy behavior.
\begin{figure}[ht]
\centerline{\includegraphics*[scale=0.4]{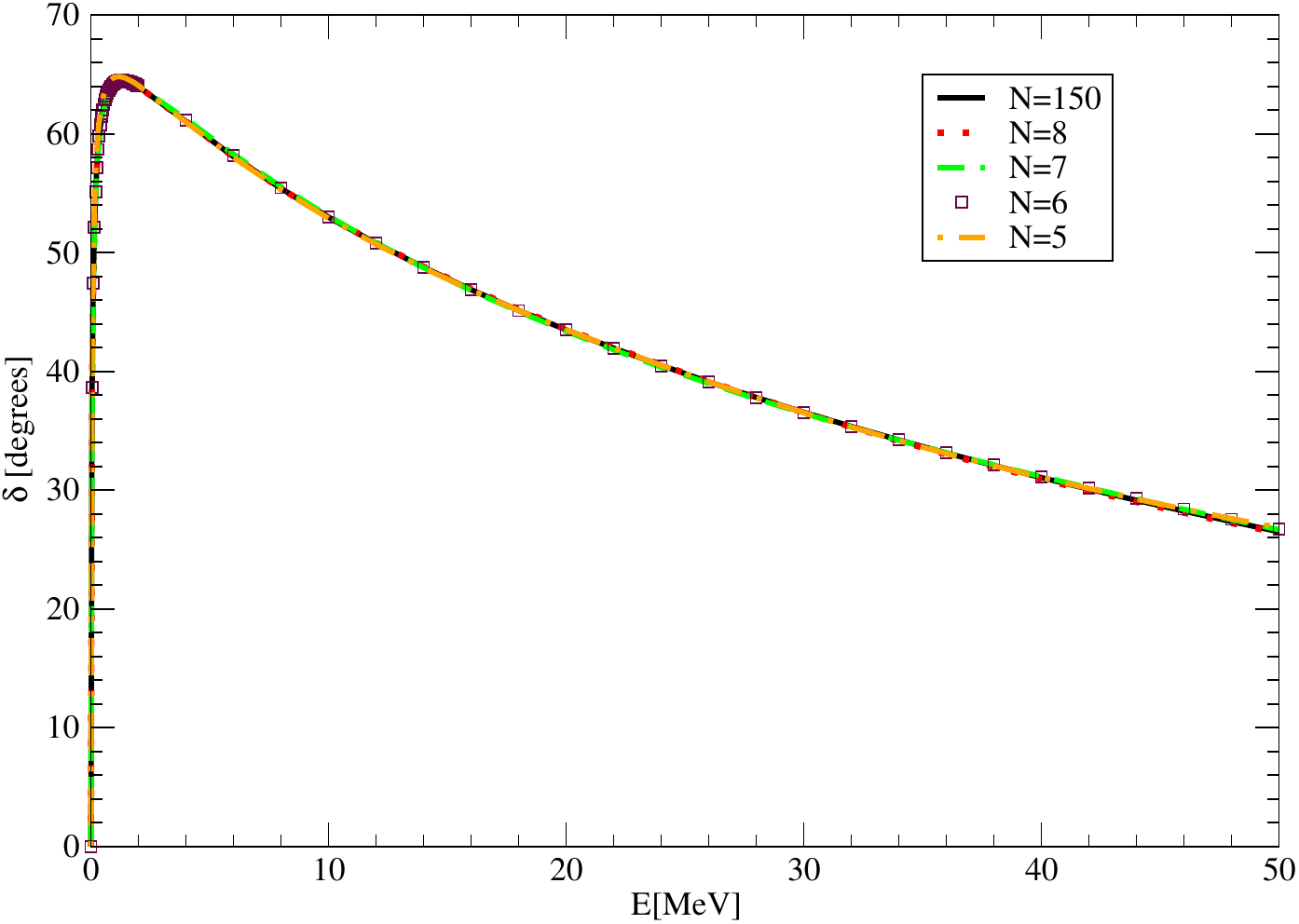}}
\centerline{\includegraphics*[scale=0.4]{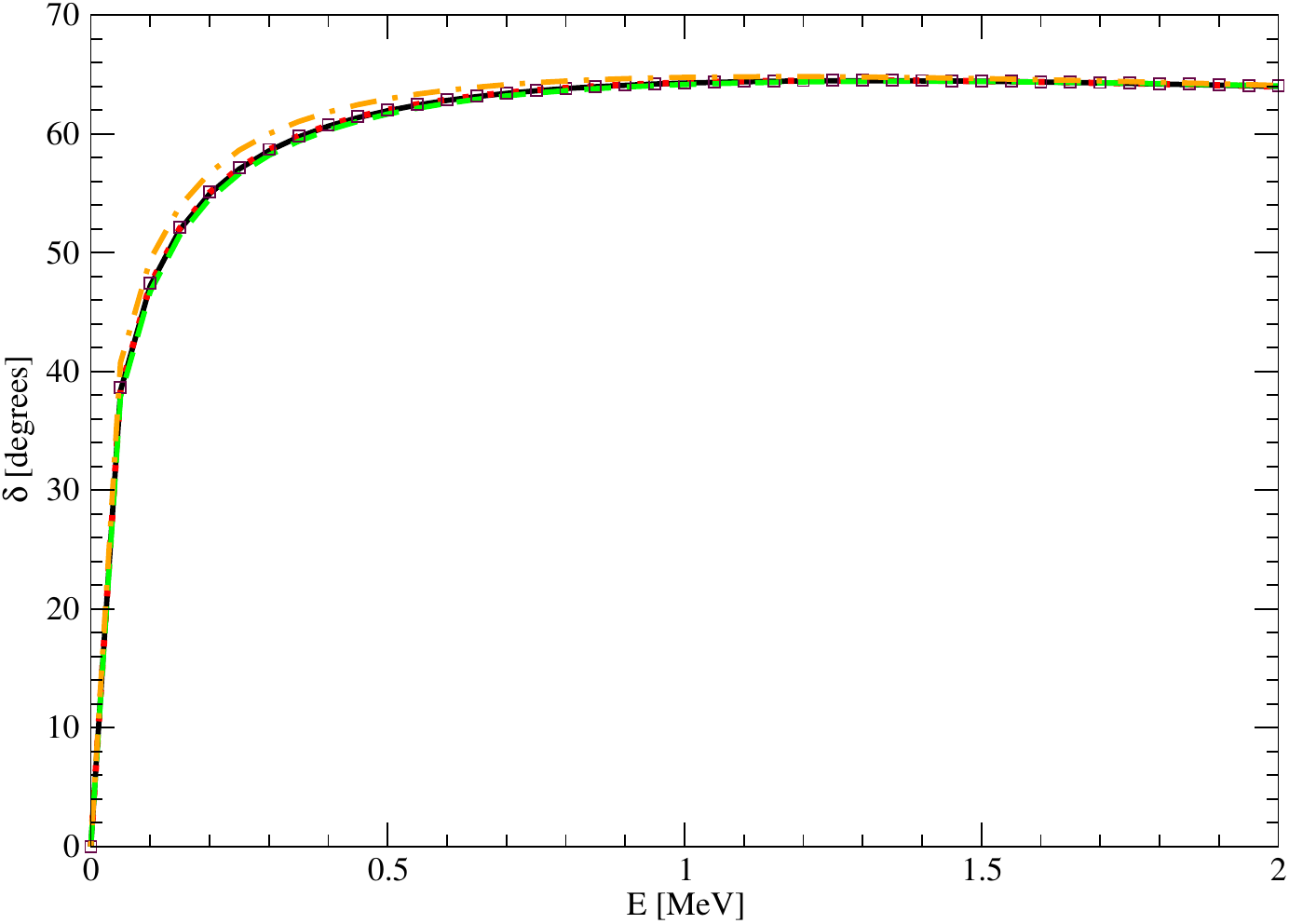}}
\caption{$^1S_0$ phase shift $\delta(E)$ computed using the regularizing factor $g_0(r)=\tanh(r/r_{cut}^0)$
with $N$  projection equations retained.  
The bottom panel highlights low-energy behavior.}
\end{figure}
One sees that excellent convergence is 
attained although only LF with low numbers of excitation quanta are retained in Eq. (\re{psi}). 
Noticeable devations from the exact phase shift appear only for $N=5$ at low energies.
Table 1 represents deviations from the exact phase shift more precisely.
\begin{table}[h]
\caption{Dependence of the phase shift $\delta(E)$ on the
number $N$ of the projection equations retained in the calculation 
with the choice $g_0(r)=\tanh(r/r_{cut}^0)$. The phase shift is measured in degrees, and
the energy~$E$ is given in MeV.}
\vspace{0.2cm}
\begin{tabular}{|c|l|l|l|l|}
\hline
$N$&$\delta(0.05)$&$\delta(2)$&$\delta(24)$&$\delta(50)$\\
\hline
5&40.8&64.08&40.524&26.9\\
\hline
6&38.7&64.030&40.475&26.7\\
\hline
7&37.89&64.040&40.36&26.7\\
\hline
8&38.50&63.979&40.561&26.39\\
\hline
9&38.9&64.009&40.5075&26.490\\
\hline
10&38.25&63.973&40.5107&26.4956\\
\hline
11&38.25&63.9926&40.521&26.475\\
\hline
50&38.37333656&63.98909667&40.506527914&26.4989739384\\
\hline
100&38.37333641410&63.98909664491&40.506527891865&26.4989739361807\\
\hline
150&38.37333641410&63.98909664491&40.506527891864&26.4989739361803\\
\hline
\end{tabular}\\
\end{table}

The results corresponding to the choice (\re{fact2}): $g_0(r)=[1-\exp(-r/r_{cut}^0)]^3$ are 
summarized below. The minimization
of the quantity (\re{opt}) is performed at $N=11$. The resulting optimal
$r_{cut}^0$ values are found to be 1.75 fm at $E=10$ and 24 MeV, while they are 2 fm for all other energies.  
The value $r_{cut}^0=2$~fm for all the $E$ and $N$ values is adopted in the subsequent calculations.
 (It is checked directly that this choice, indeed, provides the "optimal"
phase shifts.) Fig. 2 shows the results obtained. 
\begin{figure}[ht]
\centerline{\includegraphics*[scale=0.4]{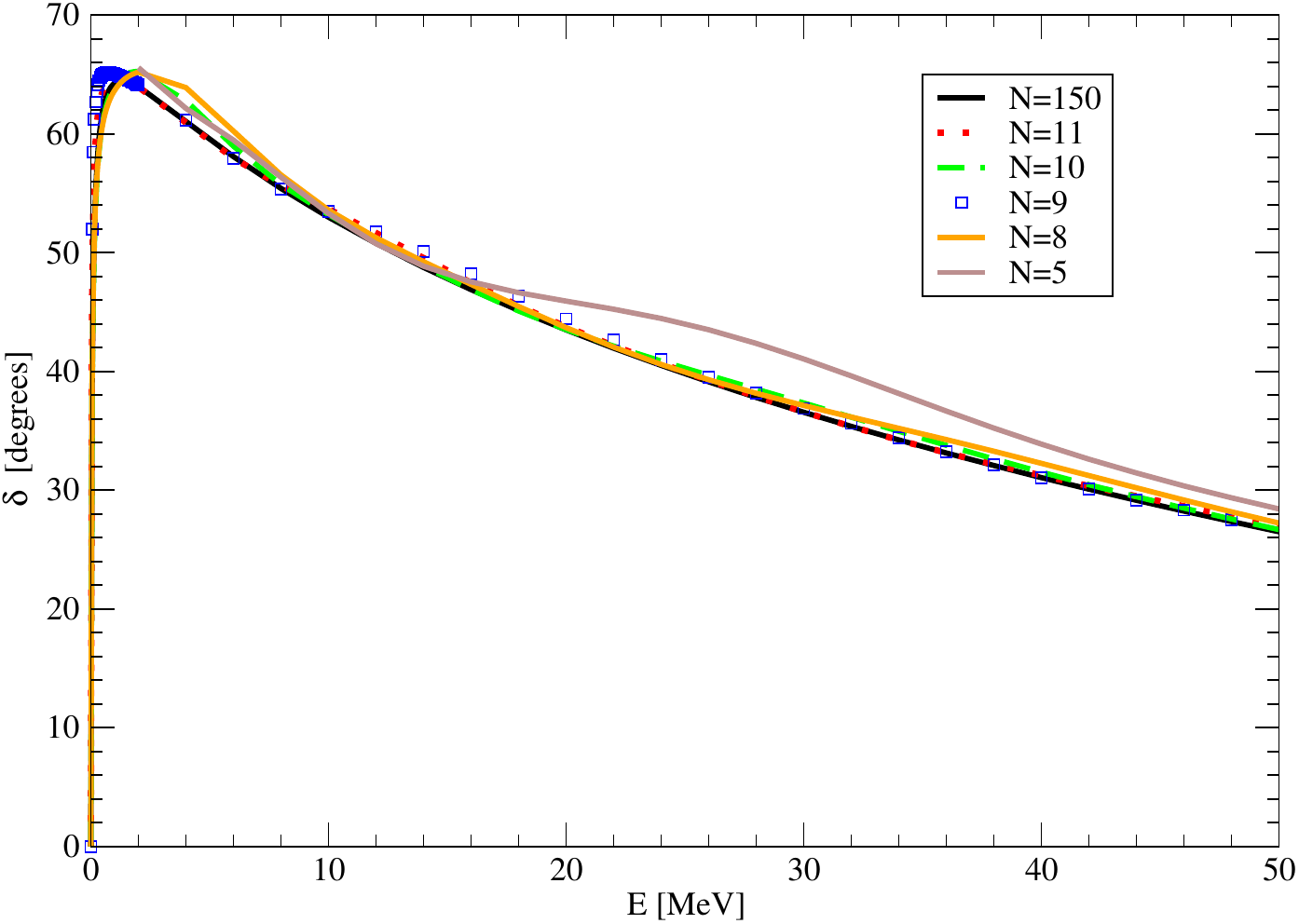}}
\caption{$^1S_0$ phase shift $\delta(E)$ computed 
using the regularizing factor $g_0=[1-\exp(-r/r_{cut}^0)]^3$  with $N$ projection equations  retained. 
The $N=5$ results are shown for
energies higher than 2 MeV.}
\end{figure}

It is seen that in this case the convergence behavior is also stable, although it proceeds at a
slower rate compared to the preceding case.
The deviations between the exact phase shift and the approximate phase shifts computed with $N$ ranging from 6 to 10
 are particularly pronounced in the low-energy regime. When $N$ reaches 11, practical convergence is achieved for energies
 higher than a few tenths of MeV. Table 2 illustrates that at very low energy of 
$E=0.05$~MeV convergence is quite gradual.

\begin{table}[h]
\caption{Dependence of the phase shift $\delta(0.05)$ for energy $E=0.05$ MeV on the
number $N$ of the projection equations retained in the calculation with 
the choice $g_0(r)=[1-\exp(-r/r_{cut}^0)]^3$. The phase shift is measured in degrees.}
\vspace{0.2cm}
\begin{tabular}{|c|c|c|c|c|c|}
\hline
$N$&9&11&50&100&150\\
\hline
$\delta(0.05)$&52&46&38.12&38.311&38.349\\
\hline
\end{tabular}\\
\end{table}

The reason for slow convergence in this case is, probably, the following. 
Denote the scattering length as $a$, $a\simeq-23.7$ fm, and the range of the potential as $R$. 
At low energy, for such a choice of $g_0(r)$, the contribution
$\tan\delta\cdot g_0(r)\cos(kr)/kr$ varies inside the potential from the large value of about $-a/R$ to zero, whereas the
true wave function varies little inside the potential. Therefore, to reproduce the true wave function, the LFs $\chi_k$
must compensate for the decline of this contribution. 
However, if their number is not large, for this purpose they need to have amplitudes on the scale of $-a/R$ which
deteriorates the description of wave function beyond the potential. This incompatibility in describing the wave
function both inside and beyond the potential can be overcome only when the number of LFs is sufficiently large.

In Refs. \cite{rev1,rev2}
 the following procedure to improve convergence in bound-state and scattering problems has been suggested.
Suppose that matrix elements of interaction are denoted as $V_{nn'}$, where the subscripts 
$n$ and $n'$ enumerate basis LF.
Let  the net 
number of basis LFs retained in a calculation be $N$.
One may modify the matrix elements $V_{nn'}$ replacing them with $V_{nn'}f(n,N)f(n',N)$, where the function $f$
is such that $f(n,N)\rightarrow 1$ when $N\rightarrow\infty$ at fixed $n$. The initial problem and the
problem with the modified matrix elements  have the same solution  in the $N\rightarrow\infty$ limit.  One may impose
the condition that the function $f(n,N)$ vanishes 
when $n=N$. In this case, the procedure  smoothes the truncation of a potential
at $n=N$ or $n'=N$.
Otherwise, the function $f$ is quite arbitrary and one may try to choose it such that the rate of convergence with $N$
of binding
energies or phase shifts  increases. Such a procedure is known in mathematics. 
In Refs.  \cite{rev1,rev2} a specific choice of the function $f$ has been
 made and an improvement of convergence has been demonstrated.
The procedure was applied in Refs. \cite{sha,sha1} 
using the aforementioned choice of the function~$f$. In our case, a more
direct and natural way to improve convergence would be a proper modification of the relative-motion functions, i.e., of
the $g_l$ factors. Therefore, we do not try using this procedure.\footnote{The results of 
Refs.~\cite{sha,sha1} without this procedure are presented in Fig. S8 of 
Ref.~\cite{sha1}. They are not sufficiently stable.}

\subsection{Two-channel scattering}

Here, scattering in the coupled $^3S_1$-$^3\!D_1$ partial waves is considered. 
In Eq. (\re{psi}), we have $Q=2$, $q=1$ and 2, and 
$i=1$ or 2. We set 
\ber\psi^{(0)}_1=j_0(kr){\cal Y}_{1M}^{01},\quad  \psi^{(0)}_2=j_2(kr){\cal Y}_{1M}^{21},\nonumber\\
\psi^{(1)}_1=g_0(r)n_0(kr){\cal Y}_{1M}^{01},\quad \psi^{(1)}_2=g_2(r)n_2(kr){\cal Y}_{1M}^{21}.\la{f}\eer
The factors $g_l(r)$ are taken as in Eq.~(\re{gl}). One has $c^i_k=K_{qi}$, where 
$K_{qi}(E)$ is the element of $K$~matrix and $q=k-P$. 
Here we use the conventional definition of $K$~matrix, such as that, e.g., in Refs. \cite{sha,al}.
It differs from the definition \cite{baz} used in Ref. \cite{efr}. In terms of the present definition, 
the $K$ matrix of Ref.~\cite{efr} is $-K^{-1}$. With the present definition, $K$ matrix
is regular at zero energy. 

The $K$ matrix is calculated from Eqs. (\re{eqs}).  While for the exact $K$ matrix one has \mbox{$K_{12}=K_{21}$},
for the calculated $K$ matrix this relation holds only approximately. The off-diagonal elements of the $K$ matrix
are replaced with their average value
\mbox{$(K_{12}+K_{21})/2$}.

The $S$ matrix is obtained from the calculated $K$ matrix:
\be S=(iI-K)(iI+K)^{-1}.\la{s}\ee
This $S$ matrix
is exactly unitary and symmetric. The phase shifts $\delta_0(E)$ and
$\delta_2(E)$, as well as the mixing parameter $\epsilon(E)$ that enter the Stapp parametrization 
\[ S(E)=\left(\begin{array}{cc}
         e^{2i\delta_0}\cos2\epsilon&ie^{i(\delta_0+\delta_2)}\sin2\epsilon\\
         ie^{i(\delta_0+\delta_2)}\sin2\epsilon&e^{2i\delta_2}\cos2\epsilon\end{array}\right)\]
are extracted from this $S$ matrix.

The localized functions $\chi_k$ from Eqs. (\re{psi})  
have the form (\re{bas}) with $J=S=1$ and $l=0$ and 2. In the calculations, functions $\chi_k$ with $l=0$ and
$n$ up to some $n_{max}$ and with $l=2$ and $n$ up to $n_{max}-1$ are retained so that the maximum oscillator 
excitation is $2n_{max}$. Eqs.~(\re{eqs}) include functions ${\bar\chi}_k$ such that ${\bar\chi}_k=\chi_k$ augmented with
two extra functions ${\bar\chi}_k$ 
with next $n$ values - one with $l=0$ and $n=n_{max}+1$ and one with $l=2$ and $n=n_{max}$. 

The number
of retained equations~(\re{eqs}) with ${\bar\chi}_k$ having $l=0$ will be denoted by $N$. 
One has $N=n_{max}+2$ so that 
the total number of retained
equations is $2N-1$. The maximum oscillator excitation is $2N-4$.

At moderate $N$ values, the results are rather sensitive 
to the choice of the parameters $r_{cut}^0$ and $r_{cut}^2$ 
entering the $g_0$ and $g_2$ regularizing factors  from Eq.~({\re f}).
The "optimal" values of these parameters 
are found as follows. A grid $r_{cut}^0=0.25\,{\rm fm}\lcdot i$ and $r_{cut}^2=0.25\,{\rm fm}\lcdot j$ is 
employed. Let us 
denote the $K$ matrix element $K_{\alpha\beta}(r_{cut}^0,r_{cut}^2)$ as $f(i,j)$. Let us use the notation 
\[f_i(i,j)=[f(i+1,j)-f(i-1,j)]/f(i,j),\quad f_j(i,j)=[f(i,j+1)-f(i,j-1)]/f(i,j),\]
and $\Delta(i,j)=[f_i^2(i,j)+f_j^2(i,j)]^{1/2}$. The values of $r_{cut}^0$ and $r_{cut}^2$  that
minimize  $\Delta(i,j)$ in the range 1 fm$\le r_{cut}^0,r_{cut}^2\le$ 3.75 fm are sought. 
The minimizations are performed at $N=11$. 

For the choice $[\tanh(r/r_{cut}^l)]^{2l+1}$ of the $g_l$ factors,
the search for the "optimal" $r_{cut}^0$ and $r_{cut}^2$ parameters yields the following results. 
At an energy of $E=2$~MeV,
the values obtained are \mbox{$(r_{cut}^0,r_{cut}^2)=(2.25, 3)$~fm} for all $K_{ij}$ matrix elements.
For $E=10$~MeV, these values are (2.5, 2.75)~fm for all $K_{ij}$ matrix elements except $K_{12}$, where
$r_{cut}^2=3$~fm. For $E=24$~MeV, $r_{cut}^0=2.5$~fm for all $K_{ij}$ matrix elements,
 while $r_{cut}^2=2.75$~fm for $K_{11}$
and $K_{12}$, and $r_{cut}^2=2.5$~fm for $K_{21}$
and $K_{22}$. For $E=50$~MeV, $r_{cut}^2=2.25$~fm for all the $K_{ij}$ matrix elements, while $r_{cut}^0=2.5$~fm
for $K_{11}$ and $K_{12}$, and $r_{cut}^0=2.75$~fm for $K_{21}$ and $K_{22}$. For all~$N$, in the subsequent
calculations  below,  a fixed value $r_{cut}^0=2.5$ fm is used for all energy values, while 
$r_{cut}^2(E)$ varies linearly between $r_{cut}^2=3$~fm at $E=2$ MeV and $r_{cut}^2=2.25$~fm at $E=50$~MeV.

The results obtained in this case are presented in Fig. 3. 

\begin{figure}[ht]
\centerline{\includegraphics*[scale=0.4]{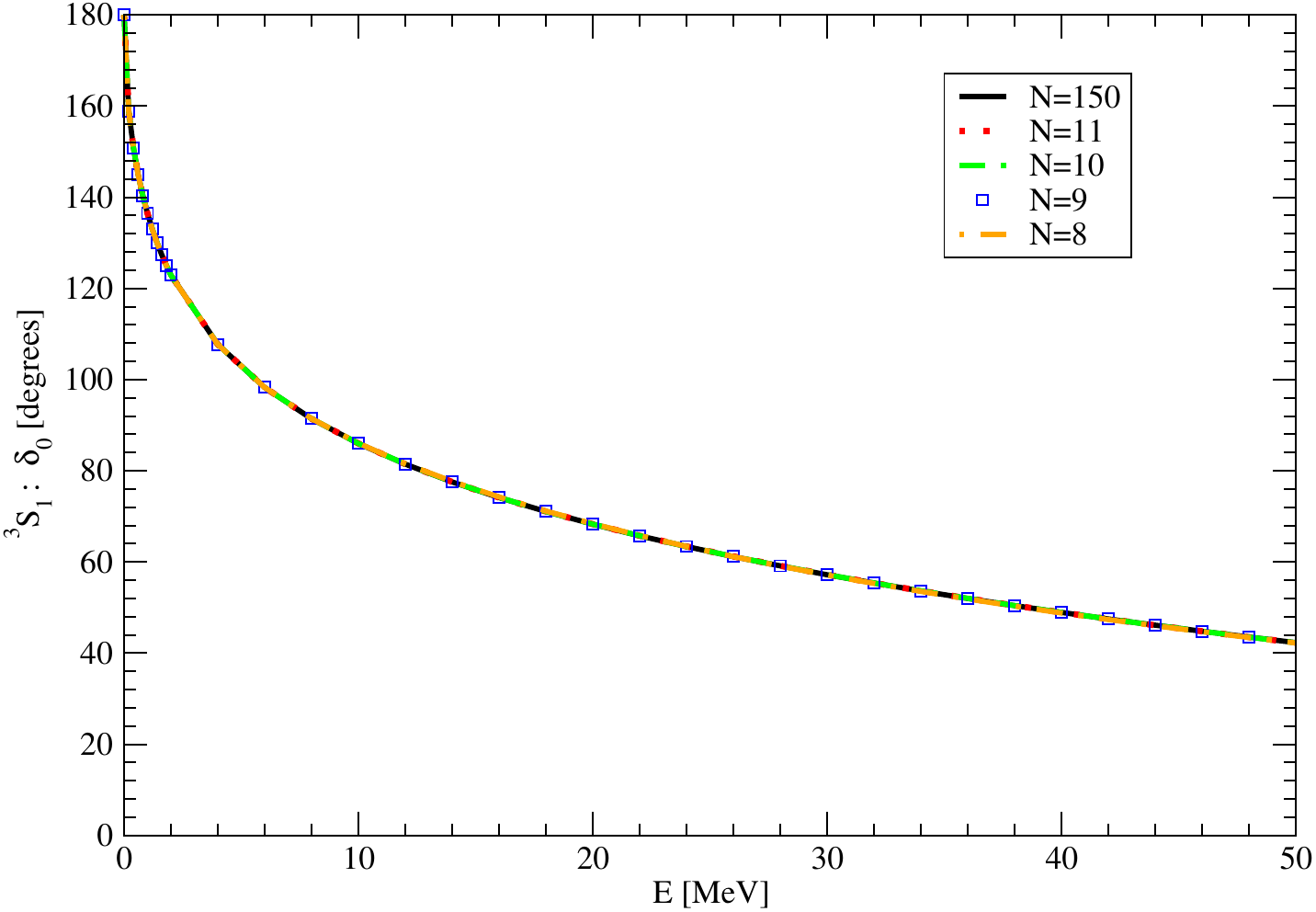}}
\centerline{\includegraphics*[scale=0.4]{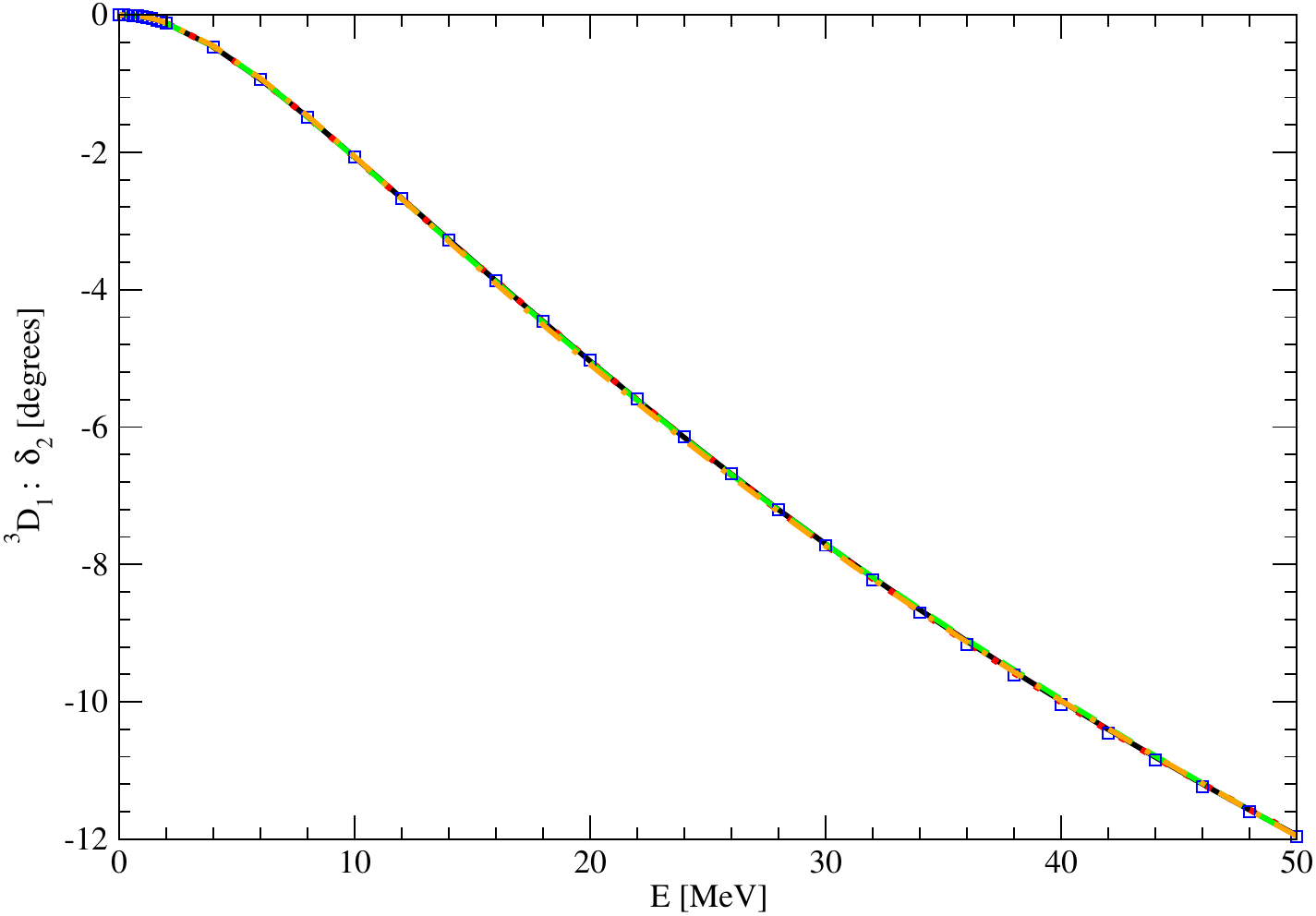}}
\centerline{\includegraphics*[scale=0.4]{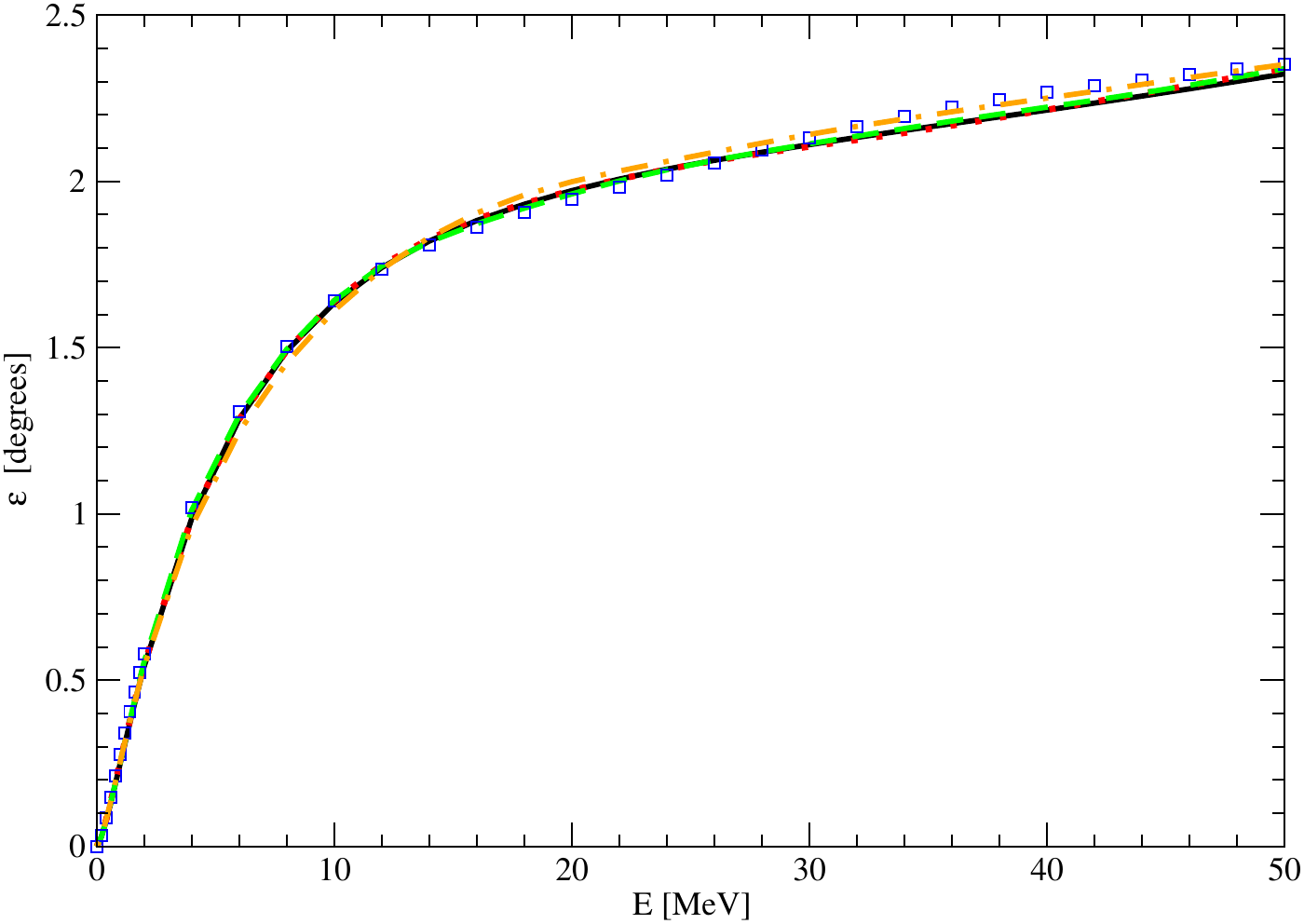}}
\caption{Scattering parameters  calculated for the choice $[\tanh(r/r_{cut}^l)]^{2l+1}$ of the regularizing factors,
when $2N-1$ 
 equations (\re{eqs}) are retained. 
Upper panel: the $^3S_1$ phase shift $\delta_0(E)$. Middle panel: the
$^3D_1$ phase shift~$\delta_2(E)$. Bottom panel: the mixing parameter $\epsilon(E)$.}
\end{figure}


It is seen that convergence of the results is very stable and it is
excelent for the phase shifts $\delta_0$ and $\delta_2$. For the mixing parameter, 
practical convergence is  achieved as well at $N$=10 and 11. 
At energies below 15 MeV, practical convergence is also observed for $N=9$, and nearly so for $N=8$.
For higher energies,
the deviations of the mixing parameter from the true values
   are still rather small at $N=8$ and 9.  

For the choice $[1-\exp(-r/r_{cut}^l)]^{2l+3}$ of the $g_l$ factors, the final
calculations  below have been performed using
the same $r_{cut}^0=2$~fm  and $r_{cut}^2=1.5$~fm values for all energies $E$ and all $N$. These  
values are determined through the search procedure listed above. However, it is found that now
at an energy of 2 MeV no required minimum exists. Nevertheless, with the listed "optimal" $r_{cut}^0$ and $r_{cut}^2$
 values, very
good convergence of the results is observed across the entire range of energies.

The results obtained in this case  are shown in Fig. 4.

\begin{figure}[ht]
\centerline{\includegraphics*[scale=0.4]{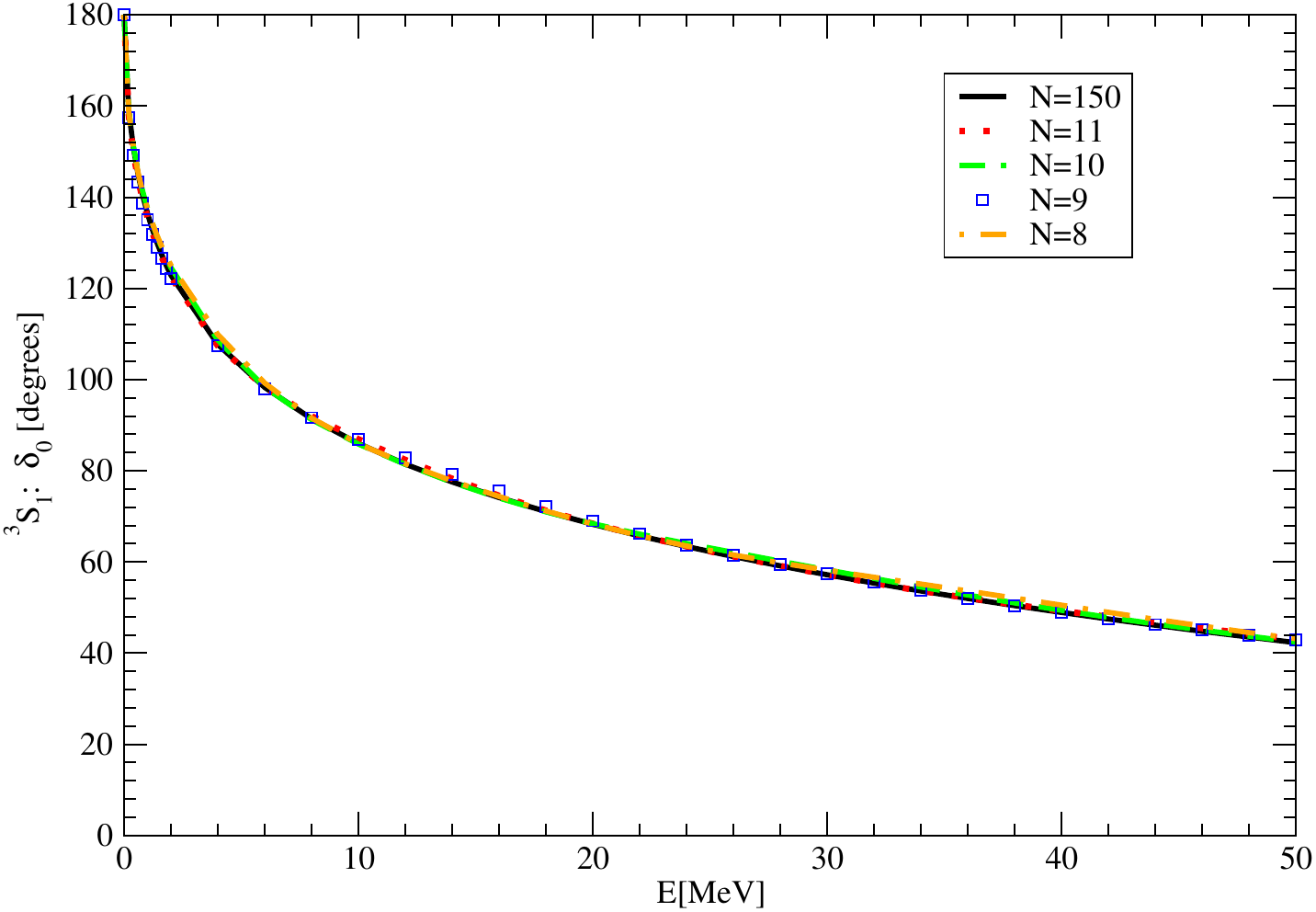}}
\centerline{\includegraphics*[scale=0.4]{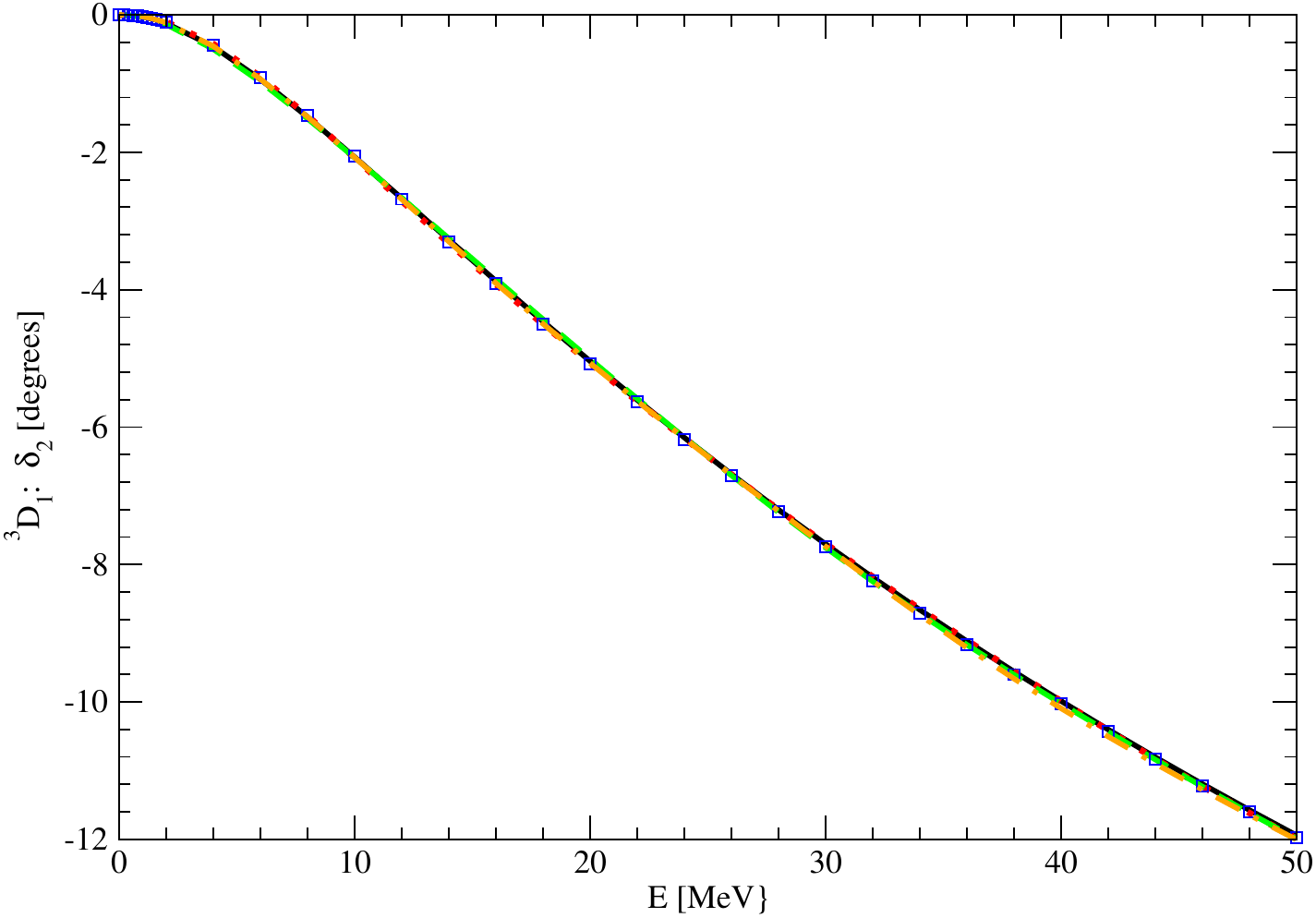}}
\centerline{\includegraphics*[scale=0.4]{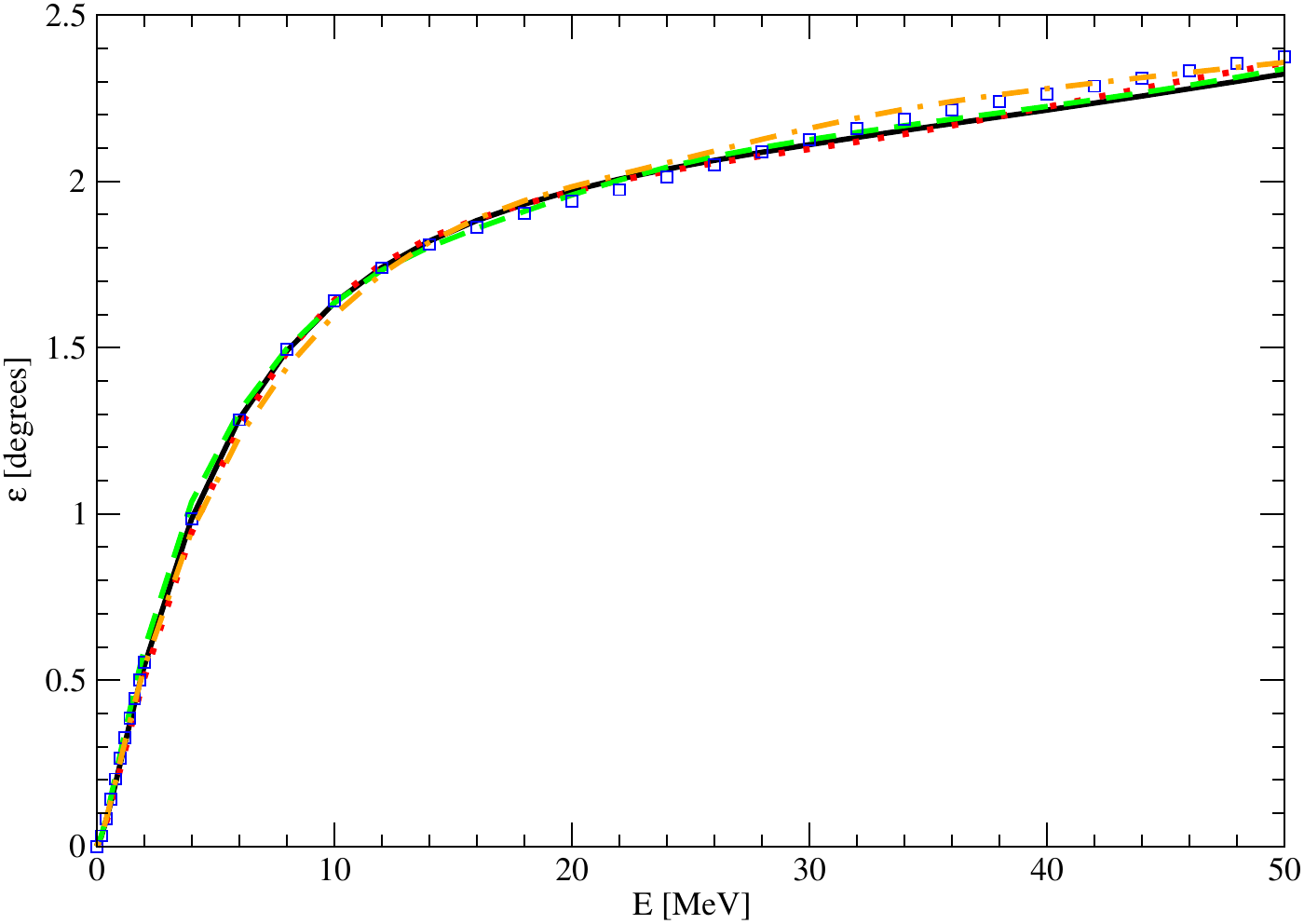}}
\caption{Same as in Fig. 3 for the choice $[1-\exp(-r/r_{cut}^l)]^{2l+3}$  of the regularizing factors.}
\end{figure}
It is seen that, in contrast to the one-channel case, here 
the rate of convergence with the chosen $g_l$ factors is nearly the same as with $g_l$ factors chosen above.

The results presented in this section, 
along with those of Ref. \cite{efr} for many-body matrix elements, indicate that the present
method has significant potential for ab initio studies of low-energy reactions.

The author appreciates conversations with A.M. Shirokov which stimulated this study.


\clearpage

\begin{thebibliography}{99}
\bibitem{john} C.W. Johnson {\it et al.,} J. Phys. G {\bf 47} 123001 (2020).
\bibitem{sha} M.A. Sharaf, A.M. Shirokov, W. Du, and J.P. Vary, Phys. Rev. C {\bf 112}, 024517 (2025).
\bibitem{qua} S. Quaglioni and P. Navr\'atil, Phys. Rev. Lett. {\bf 101}, 092501 (2008).
\bibitem{nav} P. Navr\'atil, R. Roth, and S. Quaglioni, Phys. Rev. C {\bf 82}, 034609  (2010) and references therein.
\bibitem{gys} P. Gyspers, P. Navr\'atil, K. Kravaris, G. Hupin, and S. Quaglioni, 
Phys. Rev. C {\bf 110}, 015503 (2024) and references therein.
\bibitem{efr} V.D. Efros, Phys. Rev. C {\bf 99}, 034620 (2019).
\bibitem{efr1} V.D. Efros, W. Leidemann, and V.Yu. Shalamova, Few-Body Syst. {\bf 60}, 35 (2019).
\bibitem{sha1} M.A. Sharaf, A.M. Shirokov, W. Du, and J.P. Vary, Supplemental Material to 
Ref. \cite{sha}: http://link.aps.org/supplemental/10.1103/bhgh-d9s7. These results are also presented in the
e-print by the authors, arXiv: 2408.05656, v.1. 
\bibitem{da} A.M. Shirokov, I.J. Shin, Y. Kim, M. Sosonkina, P. Maris, and J.P. Vary, Phys. Lett. B {\bf 761}, 87 (2016).
\bibitem{zhuefr} M.V. Zhukov and V.D. \'Efros, Yad. Fiz. {\bf 14}, 577 (1971) [Sov. J.
Nucl. Phys. {\bf 14}, 322 (1972)].
\bibitem{efri} V.D. Efros, Indian J. Phys. {\bf 100}, 1561 (2026).
\bibitem{dem} Yu.N. Demkov, {\it Variational Principles in the Theory of Collisions} (Macmillan, N.Y., 1963; 
Pergamon, Oxford, 1963).
\bibitem{sch} C. Schwartz, Ann. Phys. {\bf 16}, 36 (1961).
\bibitem{nut} J. Nuttall, Ann. Phys. {\bf 52}, 428 (1969).
\bibitem{rev1} B. Gyarmati, A.T. Kruppa, and J. R\'evai, Nucl. Phys. A {\bf 326} 119 (1979).
\bibitem{rev2} J. R\'evai, M. Sotona, and J. \^Zofka, J. Phys. G {\bf 11}, 745 (1985).
\bibitem{al} Y. Alhassid, G.F. Bertsch, and P. Fauto, Ann. Phys. {\bf 419}, 168233 (2020).
\bibitem{baz} A.I. Baz', Y.B. Zeldovich, and A.M. Perelomov, {\it Scattering, Reactions and Decays in Nonrelativistic
Quantum Mechanics} (Nauka, Moscow, 1971, 2nd ed.; Israel Program Sci. Transl., Jerusalem, 1969, 1st ed.)  
\end{thebibliography}
\end{document}